\documentclass[aoas,MSNbibl,nameyear,dvips]{arximspdf}
\usepackage{dcolumn}
\usepackage{multirow}
\usepackage{graphicx}

\doi{10.1214/12-AOAS552} 
\volume{6}
\issue{3}
\pubyear{2012}
\firstpage{1185}
\lastpage{1208}

\makeatletter

\setattribute{copyright}{owner}{In the public domain}

\newcolumntype{d}[1]{D{.}{.}{#1}}

\newproclaim{remark}{Remark}

\renewcommand{\mid}{|}

\newcommand{\bfbeta}{{\bolds\beta}}
\newcommand{\bfmu}{{\bolds\mu}}
\newcommand{\bfSigma}{{\bolds\Sigma}}
\newcommand{\bftheta}{{\bolds\theta}}
\newcommand{\bfzeta}{{\bolds\zeta}}

\makeatother

\begin{document}
\begin{frontmatter}

\title{Semiparametric regression in testicular germ cell~data}
\runtitle{Semiparametric regression}

\begin{aug}
\author[A]{\fnms{Anastasia} \snm{Voulgaraki}\thanksref{t1}\ead[label=e1]{avoulgar@math.umd.edu}},
\author[A]{\fnms{Benjamin} \snm{Kedem}\corref{}\thanksref{t1}\ead[label=e2]{bnk@math.umd.edu}}
\and~%
\author[B]{\fnms{Barry~I.}~\snm{Graubard}\ead[label=e3]{graubarb@exchange.nih.gov}}
\runauthor{A. Voulgaraki, B. Kedem and B. I. Graubard}
\affiliation{University of Maryland, University of Maryland and~National~Cancer~Institute}
\address[A]{A. Voulgaraki\\
B. Kedem\\
Department of Mathematics\\
University of Maryland\\
College Park, Maryland 20742\\
USA\\
\printead{e1}\\
\hphantom{E-mail: }\printead*{e2}} 
\address[B]{B. I. Graubard\\
National Cancer Institute\\
Rockville, Maryland 20852\\
USA\\
\printead{e3}}
\end{aug}

\thankstext{t1}{Supported by NSF Grant DMS-10-07647.}

\received{\smonth{9} \syear{2011}}
\revised{\smonth{2} \syear{2012}}

%
\begin{abstract}
It is possible to approach regression analysis with random
covariates from a semiparametric perspective where information
is combined from multiple multivariate sources.
The approach assumes a~sem\-iparametric density ratio model
where multivariate distributions are ``regressed'' on a reference
distribution. A kernel density estimator can be constructed from
many data sources in conjunction with
the semiparametric model.
The estimator is shown to be more efficient than the traditional
single-sample kernel density estimator, and its optimal bandwidth is
discussed in some detail. Each multivariate distribution and the
corresponding conditional expectation (regression) of interest are
estimated from the combined data using all sources. Graphical and
quantitative diagnostic tools are suggested to assess model validity.
The method is applied in quantifying the effect of height and age on
weight of germ cell testicular cancer patients. Comparisons are made
with multiple regression, generalized additive models (GAM) and
nonparametric kernel regression.
\end{abstract}

%

\end{frontmatter}

\section{Introduction}

This paper addresses the relationship between weight, height and age of
germ cell testicular cancer patients. The problem is approached by a
nonlinear regression method based on the so-called density ratio model.
The method fuses or combines information from multiple sources in order
to create an efficient kernel density estimator, which is then used in
the direct estimation of the conditional expectation, bypassing
linearity and the normal assumption. The choice of bandwidth parameters
associated with the density kernel estimates is discussed in some
detail.




In Section~\ref{StatisticalFormulation} we present the general
multidimensional semiparametric density ratio model, review the
procedure for estimating the distributions and parameters of the
model,
and discuss the asymptotic behavior of the estimators. In
Section~\ref{Combinedsemiparametricdensityestimators} we introduce the
combined (from many samples) semiparametric multivariate kernel density
estimator, and show that it is more efficient than the traditional
single-sample kernel estimator. Moreover, we discuss the associated
problem of bandwidth selection. Section~\ref{Semiparametricregression}
deals with a~semiparametric approach to regression with random
covariates, that is, semiparametric estimation of $E[y| \mathbf{x}]$.
The proposed estimator of $E[y| \mathbf{x}]$ may be viewed as a
semiparametric extension of the Nadaraya--Watson nonparametric
estimator.
We also propose various diagnostic measures to check model validity.
The method is first
illustrated by a simulation study in
Section~\ref{SomeSimulationResults} and is then applied
in Section~\ref{ApplicationtoTesticularGermCellCancer}
to Testicular Germ Cell Tumor (TGCT) data.
A comparison with other methods is made in both Sections~\ref{SomeSimulationResults} and
\ref{ApplicationtoTesticularGermCellCancer}.

\subsection{Motivation}


The $p$-dimensional formulation of the model was motivated by an
extension of a previous analysis of two risk factors, body weight and
height, of germ cell testicular cancer to including three or more risk
factors or covariates; see \citet{Kedetal09}.
Increased height has been shown to be associated with increased risk of
germ cell testicular cancer in a majority of studies, reflecting
exposure to, possibly, early life factors due to genetics, nutrition or
endogenous or exogenous hormones; see \citet{McGCoo10}.
Body weight reflects potentially later life exposures such as dietary
intake and energy expenditure behavior. A few studies have found that
increased body mass (body weight divided by height squared) was
associated with a decrease in risk of testicular cancer, but most
studies have found no association [\citet{McGCoo10}].
This lack of association may be due to inappropriate parametric
modeling, usually logistic regression. The use of a two-dimensional
density ratio model in the previous analysis uncovered an important
contribution of body weight in the presence of height that was not
observed in logistic regression analyses; see \citet{McGetal07}.
We wanted to include age in the analysis with height and weight as age
is both an important risk factor and potential confounder since the
incidence of testicular cancer varies by age, peaking around 25--35
years for the most common types of testicular cancer, and age
correlates with body weight; see \citet{McGCoo10} and
\citet{Ogdetal04}.
The proposed extension of the density ratio model provides an
opportunity to explore the interrelationships of height and weight with
testicular cancer while controlling for age by estimating the
conditional expectation of weight given height and age.


\subsection{Background and preliminaries}

Suppose there are $m=q+1$ data sources, such as $q$ case groups and a
control group,
each giving a sample of random vectors from an
unknown multivariate distribution.
In the density ratio model one of these distributions serves as a
reference\vadjust{\goodbreak} or baseline, and all other distributions are tilts of the
reference. In its one-dimensional form the model is motivated by the
classical one-way analysis of variance with \mbox{$m=q+1$} independent normal
random samples, and logistic regression; see
\citet{Foketal01} and \citet{QinZha97}.
In its multivariate form, the model is motivated by classical
classification given multivariate normal samples, and generalized
logistic regression; see \citet{And71} and \citet{PrePyk79}.

In the one-dimensional case there are $m=q+1$ random samples,
\[
(x_{11},\ldots,x_{1n_1}), \ldots,
(x_{q1},\ldots,x_{qn_q}), (x_{m1},\ldots,x_{mn_m})
\]
with probability density functions $g_i$,
%
\begin{equation}\label{eqmsamples}
x_{ij} \sim g_{i},\qquad i=1,\ldots,q,m, j=1,\ldots,n_i,
\end{equation}
where $ g_{m} \equiv g$ is called the \textit{reference} probability
density. Assuming exponential tilts, the $g_i$ satisfy the
(exponential) \textit{density ratio model}
%
\begin{equation}\label{eqDensityRatioModel}
\frac{g_j(x)}{g(x)} = \exp\bigl(\alpha_j +
\bfbeta^{\prime}_j \mathbf{h}(x)\bigr),\qquad j=1,\ldots,q.
\end{equation}
It is assumed that
the \textit{distortion function} $\mathbf{h}(x)$ is a known vector-valued
function. The objective is to estimate the reference density $g$,
the corresponding cumulative distribution function (CDF) $G$ and
the parameters~$\alpha_j,\beta_j$ from the combined data
%
\begin{equation}\label{eqCombineddatat1,,tn}
\mathbf{t}
= \{(x_{11},\ldots,x_{1n_1}), \ldots,
(x_{q1},\ldots,x_{qn_q}),
(x_{m1},\ldots,x_{mn_m})\}^\prime.
\end{equation}

The density ratio model has been
applied in various problems including
kernel density estimation
[\citet{Fok04}, \citet{CheChu04}, \citet{QinZha05}],
analysis of variance [\citet{Foketal01}],
AIDS vaccine trials
[\citet{GilLelVar99}],
mortality rate prediction [\citet{Kedetal08}],
microarrays evaluation [\citet{Phuetal07}],
case-control studies [\citet{PrePyk79}, \citet{Qin98}],
logistic model validation [\citet{QinZha97}],
cluster detection [\citet{WenKed09}] and
goodness of fit [\citet{Zha00}].
A two-dimensional case-control application
has been made recently in \citet{Kedetal09}.

In this paper the asymptotic results for the semiparametric kernel
density estimator and the
estimation of the conditional expectation of a response given covariate
information are formulated under the general multiple sample
$p$-dimensional density ratio model.
Specifically, for each of the~$m$ data sources, we use the
$p$-dimensional density ratio
model in predicting, via the estimated conditional expectation,
the response variable given the corresponding covariate information,
and propose measures of goodness of fit and diagnostic
plots to check model validity.
A comparison with linear multiple regression, generalized additive
models (GAM) and the Nadaraya--Watson
kernel nonparametric regression is made using both
real and simulated data.\looseness=1

\section{Statistical formulation}
\label{StatisticalFormulation}

Suppose we have $m=q+1$ independent data sets or random samples
of
$p$-dimensional vectors $\mathbf{x}=\mathbf{x}_{p\times
1}=(x_1,x_2,\ldots, x_p)'$. Let\vadjust{\goodbreak} $g_i(x_1,x_2,\ldots, x_p)$ be the
probability function
corresponding to the $i$th sample. Assume that the $i$th sample
size is $n_i$ and $n=\sum_{i=1}^m n_i$ is the total sample size.
Thus, for $i=1,\ldots,q,m, j=1,\ldots,n_i$, we have that
\[
\mathbf{x}_{ij}=(x_{ij1}, x_{ij2},
\ldots, x_{ijp} )\sim g_{i}(x_1,\ldots,x_p)
\]
and
\[
\mathbf{x}_{i1}, \mathbf{x}_{i2},\ldots,
\mathbf{x}_{in_i} \stackrel{\mathrm{i.i.d.}}{\sim} g_{i},
\]
where $\mathbf{x}_{ij},\mathbf{x}_{ij'} $ are independent for
$j\neq j'$ and $\mathbf{x}_{ij},\mathbf{x}_{i'k} $ are independent for
$i\neq i'$ and all $j$ and $k$.
We choose $\mathbf{x}_{mj}$ as the reference sample.
Then $g\equiv g_{m}(\mathbf{x}) \equiv g_{m}(x_1,\ldots,x_p)$
is called the reference or baseline probability density function
(p.d.f.). We assume that the $g_{i}(\mathbf{x}), i=1,\ldots, q$,
satisfy the
(general) \textit{density ratio model}:
%
\begin{equation}\label{SDRM1}
\frac{g_{i}(\mathbf{x})}{g_m(\mathbf{x})}=w(\mathbf{x},\bftheta_i)
\end{equation}
or, equivalently,
%
\begin{equation}
g_{i}(\mathbf{x})=w(\mathbf{x},\bftheta_i)g_m(\mathbf{x}),
\end{equation}
where $g_{i}(\mathbf{x})$, $g_{m}(\mathbf{x})$ are not specified,
$w$ is a known positive and continuous function, and the $\bftheta_i$ are
unknown $d$-dimensional vectors of parameters.
This construction accommodates both continuous and discrete
distributions, and it does not require symmetry, let alone normality
in the continuous case.

Let $G(\mathbf{x})\equiv G_m(\mathbf{x})$ denote the reference cdf and
define $p_{ij}=dG(\mathbf{x}_{ij})=dG_m(\mathbf{x}_{ij})$. Using the
method of constrained empirical likelihood, we can estimate $g_i$ and
$\bftheta_i$ from the entire combined data, and not just from the
corresponding samples $\mathbf{x}_{ij}$ and $\mathbf{x}_{mj}$; see
\citet{Fok04}.
The empirical likelihood based on the pooled
data $\mathbf{x}_{ij}, i=1,\ldots,m, j=1,\ldots,n_i, $ is
%
\begin{eqnarray}
L(\bftheta,G_m)&=&\Biggl[\prod_{j=1}^{n_1}p_{1j}w(\mathbf{x}_{1j},
\bftheta_1)\Biggr]\Biggl[\prod_{j=1}^{n_2}p_{1j}w(\mathbf{x}_{2j},
\bftheta_2)\Biggr]
\cdots\Biggl[\prod_{j=1}^{n_m}p_{mj}\Biggr]
\nonumber\\[-8pt]\\[-8pt]
&=&\Biggl[\prod_{i=1}^m\prod_{j=1}^{n_i}p_{ij}\Biggr]
\Biggl[\prod_{i=1}^q\prod_{j=1}^{n_i}w(\mathbf{x}_{ij},\bftheta
_i)\Biggr].\nonumber
\end{eqnarray}
Let $\bftheta=(\bftheta_1',\ldots,\bftheta_q')'$, a vector of
dimension
of $qd$. The log-likelihood is given by
%
\begin{equation}\label{loglik}
l=\log L=\sum_{i=1}^m\sum_{j=1}^{n_i}\log(p_{ij})+
\sum_{i=1}^q\sum_{j=1}^{n_i}\log(w(\mathbf{x}_{ij},\bftheta_i))
\end{equation}
and is subject to the constraints
%
\begin{eqnarray}\label{generalconstraints}
p_{ij}\geq0,\qquad \sum_{i=1}^m\sum_{j=1}^{n_i}p_{ij}=1,\qquad
\sum_{i=1}^m\sum_{j=1}^{n_i}p_{ij}w(\mathbf{x}_{ij},
\bftheta_k)=1 \nonumber\\[-8pt]\\[-8pt]
&&\eqntext{\mbox{for } k=1,\ldots,q.}
\end{eqnarray}

\citet{Fok04}
and \citet{QinLaw94}
gave conditions guaranteeing that, with probability approaching $1$,
there is a maximum in a small neighborhood of the true parameter
$\bftheta_0$. Define $\mu_k\equiv\frac{\lambda_k}{n}$, where
$\lambda_k$ are the Lagrange multipliers. Then, replacing
$\mu_k$ and $\bftheta_k$ by their estimators, $p_{ij}$ and~$G_m(x)$ are
estimated by
%
\begin{eqnarray}\label{hatpij}
\hat{p}_{ij}&=&\frac{1}{n}\frac{1}{1+
\sum_{k=1}^q \hat{\mu}_k [w(\mathbf{x}_{ij},\hat{\bftheta}_k)-1]},
\\
%
%
\label{hatGm}
\hat{G}_m(\mathbf{x})&=&\sum_{i=1}^m\sum_{j=1}^{n_i}\hat
{p}_{ij}I(\mathbf
{x}_{ij}\leq\mathbf{x}) \nonumber\\[-8pt]\\[-8pt]
&=& \frac{1}{n}\sum_{i=1}^m\sum_{j=1}^{n_i}
\frac
{I(\mathbf{x}_{ij}\leq\mathbf{x})}{1+\sum_{k=1}^q \hat{\mu}_k
[w(\mathbf{x}_{ij},\hat{\bftheta}_k)-1 ]},\nonumber
\end{eqnarray}
where $I(B)$ is the indicator of the event $B$, and
$I(\mathbf{x}_{ij}\leq\mathbf{x})$ is defined componentwise.
More generally, for $l=1,\ldots, m$ and
$w(\mathbf{x}_{ij},\hat{\bftheta}_m)\equiv1$,
%
\begin{eqnarray}\label{hatGl}
\hat{G}_l(\mathbf{x})&=&\sum_{i=1}^m\sum_{j=1}^{n_i}
\hat{p}_{ij}w(\mathbf{x}_{ij},\hat{\bftheta}_l)
I(\mathbf{x}_{ij}\leq\mathbf{x}) \nonumber\\[-8pt]\\[-8pt]
&=&\frac{1}{n}\sum_{i=1}^m\sum_{j=1}^{n_i}
\frac{w(\mathbf{x}_{ij},\hat{\bftheta}_l)}{1+\sum_{k=1}^q \hat
{\mu}_k
[w(\mathbf{x}_{ij},\hat{\bftheta}_k)-1 ]} I(\mathbf{x}_{ij}
\leq\mathbf{x}).
\nonumber
\end{eqnarray}

Let $\bftheta_0$ be the true value of $\bftheta$ under model (\ref{SDRM1}).
Define the sample size ratios
$\rho_i=n_i/n_m$ and set $w(\mathbf{x},\hat{\bftheta}_i)=
w_i(\mathbf{x})$ for $i=1,\ldots,m$. Then $\rho_m\equiv1$,
$w_m(\mathbf{x})\equiv1$. We assume the $\rho_i$ are positive and
finite and remain fixed as
$n\rightarrow\infty$. Let $\bfzeta$ denote the true value of $\bfmu$.
Set $\bfzeta_n=(\bfzeta_{1n},\ldots,\bfzeta_{qn})$ and
$\zeta_{ln}=n_l/n$
for $l=1,\ldots,q$. As $n\rightarrow\infty$, assume that
$\zeta_{ln}\rightarrow\zeta_l$. Then \citet{Fok04}
showed that
$\bfzeta_n\rightarrow\bfzeta$ and that under regularity conditions
$\hat{\bftheta}-\bftheta_0$ and $\hat{\bfmu}-\bfzeta$ are jointly
asymptotically normal.
The complete statement is Theorem 1
in an Appendix in \citet{VouKedGra}.

\section{Combined semiparametric density estimators}
\label{Combinedsemiparametricdensityestimators}

Fokianos (\citeyear{Fok04}),
\citet{CheChu04}
and \citet{QinZha05}
constructed a kernel-based
density estimator by smoothing the increments of $\hat{G}_i,
i=1,\ldots,m$.
\citet{Fok04} studied the statistical properties
of the proposed kernel density estimator (mean, variance) and showed that
combining data leads to more efficient kernel density estimators under
the univariate case of the general model (\ref{SDRM1}).
\citet{QinZha05}
studied semiparametric inference
for the univariate version of model
(\ref{SDRM1}) with $w(x,\alpha, \beta)=\exp(\alpha+r(x)\beta)$.
\citet{CheChu04}
studied the same special case as \citet{QinZha05}
but followed a different approach.

In this section we aim to study the corresponding asymptotic theory and
convergence properties of the proposed kernel density estimator for the
general multivariate multiple-sample case model (\ref{SDRM1}). The
estimator is shown to be more efficient
than the traditional kernel density estimator. In addition, several
methods for calculating the optimal bandwidth are discussed. Precise
statements and proofs
are given in \citet{VouKedGra}.

The traditional kernel density estimator is a convolution of the
jumps in the empirical distribution function obtained from a
single sample of size~$n$ and a kernel function taken as a symmetric
probability density function parametrized by a bandwidth
parameter [\citet{Par62}].
Specifically, the traditional kernel density
estimator of a probability density $f(\mathbf{x})$ is given by
%
\begin{equation}\label{multkernelest}
\hat{f}(\mathbf{x})=\frac{1}{nh^p_n} \sum_{i=1}^n K \biggl(\frac
{\mathbf
{x}-\mathbf{x}_i}{h_n} \biggr),
\end{equation}
where $h_{n}$ is a sequence of bandwidths such that $h_{n} \rightarrow0$
and $nh^p_{n} \rightarrow\infty$ as $n \rightarrow\infty$. The kernel
function $K(\mathbf{x})$ is defined for $p$-dimensional $\mathbf{x}$. It
is nonnegative, symmetric around $\mathbf{0}$ and satisfies
$\int_{\mathbf{R}^p} K(\mathbf{x}) \,d\mathbf{x} =1$. Under certain
conditions, $\hat{f}(\mathbf{x})$ is
a consistent estimator of $f(\mathbf{x})$
[\citet{Par62}, \citet{Sha03}].
As such,
the traditional kernel density estimator is a ``single sample''
estimator.

Using a similar idea to (\ref{multkernelest}), we use the
the probabilities $\hat{p}_{ij}$ in (\ref{hatpij}) to form kernel
estimates for the probability densities $g_l(\mathbf{x})$,
%
\begin{equation}\label{glhat}
\hat{g}_l(\mathbf{x})=\frac{1}{h_n^p}\sum_{i=1}^m\sum
_{j=1}^{n_i}\hat
{p}_{ij}\hat{w}_l(\mathbf{x}_{ij})K\biggl(\frac{\mathbf{x}-\mathbf
{x}_{ij}}{h_n}\biggr),
\end{equation}
where $h_n$ is a sequence of bandwidths such that $h_n\rightarrow0$
and $nh^p_{n} \rightarrow\infty$ as $n\rightarrow\infty$,
$w_l(\mathbf{x})\equiv w(\mathbf{x}, \bftheta_l)$, $\hat
{w}_l(\mathbf
{x})\equiv w(\mathbf{x}, \hat{\bftheta}_l)$,
and $K$ is a nonnegative kernel function that satisfies the following
requirements:
\begin{longlist}[(3)]
\item[(1)]$\int K(\mathbf{x}) \,d\mathbf{x}=1$ and $\int|K(\mathbf{x})|
\,d\mathbf{x}< \infty$;
\item[(2)]$\int\mathbf{x} K(\mathbf{x}) \,d\mathbf{x}=\mathbf{0}$ and
$\int
|\mathbf{x} K(\mathbf{x})| \,d\mathbf{x}< \infty$;
\item[(3)]$\int\mathbf{x}' \mathbf{x} K(\mathbf{x}) \,d\mathbf{x}=k_2$ and
$\int|\mathbf{x}' \mathbf{x} K(\mathbf{x})| \,d\mathbf{x}< \infty$.
\end{longlist}


It is easy to verify that $\hat{g}_l$ is a proper probability function.

\subsection{\texorpdfstring{Asymptotic results for $\hat{g}_l$}{Asymptotic results for gl}}
\label{Asymptoticresults}

To facilitate the study of $\hat{g}_l$, it is convenient to define
first $\tilde{g}_l(\mathbf{x})$:
%
\begin{equation}
\tilde{g}_l(\mathbf{x})=\frac{1}{h_n^p}\sum_{i=1}^m
\sum_{j=1}^{n_i}p_{ij}w_l(\mathbf{x}_{ij})
K\biggl(\frac{\mathbf{x}-\mathbf{x}_{ij}}{h_n}\biggr).
\end{equation}
With this device, and with the help of
Lemmas 1--4 and Theorem 2 in \citet{VouKedGra},
in Corollary 1
in there
it is shown that
\[
\sqrt{nh_n^p}\biggl(\hat{g}_l(\mathbf{x})-
g_l(\mathbf{x})-\frac{1}{2}h_n^2
\int\mathbf{u}'\,\frac{\partial^2 g_l(\mathbf{x}^*)}
{\partial\mathbf{x}\,\partial\mathbf{x}'}\mathbf{u}
K(\mathbf{u})\,d\mathbf{u} \biggr)
\stackrel{D}{\rightarrow} N(\mathbf{0},\sigma^2(\mathbf{x}))
\]
as $n\rightarrow\infty$,
where
\[
\sigma^2(\mathbf{x})=\frac{w_l(\mathbf{x})g_l(\mathbf{x})}
{\sum_{k=1}^m \zeta_k w_k(\mathbf{x})}\int K^2(\mathbf{u})
\,d\mathbf{u}
\]
for any fixed $\mathbf{x}$.

\subsection{\texorpdfstring{Comparison of $\hat{g}_l$ and the traditional $\hat{f}$}{Comparison of gl and the traditional f}}
\label{comparisonclassicalsemiparametricestimator}

In Theorem 3 in \citet{VouKedGra} we show that
as $n\rightarrow\infty$, $h_n\rightarrow0$, and $nh_n^p\rightarrow
\infty$,
\begin{eqnarray*}
\operatorname{MISE}(\hat{g}_l)&=&\frac{1}{nh_n^p}\int\frac{w_l(\mathbf
{x})g_l(\mathbf
{x})}{\sum_{k=1}^m\zeta_kw_k(\mathbf{x})}\,d\mathbf{x}\int
K^2(\mathbf
{u})\,d\mathbf{u}\\
&&{}+\frac{h_n^4}{4}\int\biggl(\int\mathbf{u}'\,\frac{\partial^2
g_l(\mathbf
{x})}{\partial\mathbf{x}\,\partial\mathbf{x}'}\mathbf{u}K(\mathbf
{u})\,d\mathbf{u}\biggr)^2 \,d\mathbf{x}\\
&&{}+o\biggl(\frac{1}{nh_n^p}
\biggr)+o(h_n^4),
\end{eqnarray*}
from which we get the optimal bandwidth $h_n^*$ given in formula (4)
in \citet{VouKedGra}.
In Theorem 4 there
it is shown that under mild conditions
$\hat{g}_l$ is more efficient (MISE) than the traditional single-sample
$\hat{f}$ for every~$l$,
as $n\rightarrow\infty$, $h_n\rightarrow0$, and $nh_n^p\rightarrow
\infty$.

\subsection{\texorpdfstring{Bandwidth selection for $\hat{g}_l$}{Bandwidth selection for gl}}\label{BWselection}

From Section~\ref{Asymptoticresults} we see that, as is the case
with
the traditional single-sample estimator, the pooled estimator
$\hat{g}_l$ also suffers from a similar bias-variance
trade-off problem where a smaller~$h_n$ reduces the bias at the expense
of the variance, whereas a larger~$h_n$ increases the bias but
reduces the variance. We discuss next practical ways for choosing
bandwidths which are optimal in some sense.

The formula for the asymptotically optimal bandwidth $h_n^*$ given in
equation (4)
in \citet{VouKedGra} is not practical since $g_l$ is not known.
In the one-dimensional case \citet{Sil86}
proposes to either use the normal density
$N(\mu, \Sigma)$, where $\mu$ and $\Sigma$ are estimated from the
data, or $\hat{f}$ to approximate $g_l$. Following \citet{Sil86},
\citet{Fok04}
and \citet{QinZha05},\vadjust{\goodbreak}
both replace $g_l$ by
$\hat{g}_l$. However, in the multidimensional setting the
computational burden is heavier and, as \citet{Sil86}
remarks, it is somewhat hazardous to
estimate
$\partial^2 g_l(\mathbf{x})/\partial\mathbf{x}\,\partial\mathbf{x}'$
by
$\partial^2
\hat{g}_l(\mathbf{x})/\partial\mathbf{x}\,\partial\mathbf{x}'$
unless very large samples are available.


The bandwidth can also be selected via \textit{cross-validation},
which minimizes, with respect to $h_n$, an estimate for the
integrated squared error (ISE):
\begin{eqnarray*}
\operatorname{ISE}(h_n)&=&\int\bigl(\hat{g}_l(\mathbf{x})
-g_l(\mathbf{x})\bigr)^2
\,d\mathbf{x}\\
&=&\int\hat{g}_l^2(\mathbf{x})
\,d\mathbf{x}-2\int\hat{g}_l(\mathbf{x})
g_l(\mathbf{x})\,d\mathbf{x}+\int g_l^2(\mathbf{x})\,d\mathbf{x}.
\end{eqnarray*}
The last term does not depend on $h_n$, so we may drop it in the
minimization of ISE. To minimize ISE, we need to rewrite the first and
second terms as functions of $h_n$ and the data.
Denote by
\[
\mathbf{t}=[\mathbf{x}_{11}',\ldots,\mathbf{x}_{1n_1}',\ldots,
\mathbf{x}_{m1}',\ldots,\mathbf{x}_{mn_m}']'_{n \times1}=
(\mathbf{t}_1',\ldots, \mathbf{t}_n')^\prime
\]
the combined data. So $\mathbf{t}$ has $n$ rows. The first term can be written
\begin{eqnarray*}
\int\hat{g}_l^2(\mathbf{x})\,d\mathbf{x}
&=&\int\biggl[\frac{1}{h_n^p}
\sum_{i=1}^m \sum_{j=1}^{n_i} \hat{p}_{ij} \hat{w}_l(\mathbf{x}_{ij})
K\biggl(\frac{\mathbf{x}-\mathbf{x}_{ij}}{h_n}\biggr)
\biggr]^2\,d\mathbf
{x}\\
&=&\frac{1}{h_n^{2p}}\int\sum_{i=1}^m \sum_{j=1}^{n_i}\sum
_{i'=1}^m \sum
_{j'=1}^{n_i}\hat{p}_{ij}\hat{w}_l(\mathbf{x}_{ij})K\biggl(\frac
{\mathbf
{x}-\mathbf{x}_{ij}}{h_n}\biggr) \\
&&\hspace*{98pt}{}\times\hat{p}_{i'j'}\hat{w}_l(\mathbf
{x}_{i'j'})K\biggl(\frac{\mathbf{x}-\mathbf{x}_{i'j'}}{h_n}
\biggr)\,d\mathbf{x}\\
&=&h_n^{-p}\sum_{i=1}^n \sum_{i'=1}^n \hat{p}(\mathbf{t}_i) \hat
{w}_l(\mathbf{t}_i)\hat{p}(\mathbf{t}_{i'})\hat{w}_l(\mathbf
{t}_{i'})\int K(\mathbf{z})K\biggl(\mathbf{z}+\frac{\mathbf
{t}_i-\mathbf
{t}_{i'}}{h_n}\biggr)\,d\mathbf{z}.
\end{eqnarray*}
For the second term notice that
$\int\hat{g}_l(\mathbf{x})g_l(\mathbf{x})\,d\mathbf{x}= E
\hat{g}_l(\mathbf{x})$. Following \citet{Sil86} and
\citet{CheChu04}, we can estimate\vspace*{1pt} $E
\hat{g}_l(\mathbf{x})$ using the leave one out estimator $\widehat{E
\hat{g}_l(\mathbf{x})}$,
\[
\widehat{E \hat{g}_l(\mathbf{x})}=\frac{1}{n_l}\sum_{i=n_1+\cdots
+n_{l-1}+1}^{n_l} \hat{g}_{l,i}(\mathbf{t}_i),
\]
where $\hat{g}_{l,i}(\mathbf{t}_i)$ is $\hat{g}_l(\mathbf{t}_i)$ with
$\mathbf{t}_i$ dropped from the combined data.
Therefore,
a nearly optimal bandwidth $h_n$ is obtained by minimizing
%
\begin{eqnarray}\label{leaveoneout1}
&&
h_n^{-p}\sum_{i=1}^n \sum_{i'=1}^n \hat{p}(\mathbf{t}_i) \hat
{w}_l(\mathbf{t}_i)\hat{p}(\mathbf{t}_{i'})\hat{w}_l(\mathbf
{t}_{i'})\int K(\mathbf{z})K\biggl(\mathbf{z}+\frac{\mathbf
{t}_i-\mathbf
{t}_{i'}}{h_n}\biggr)\,d\mathbf{z}\nonumber\\[-8pt]\\[-8pt]
&&\qquad{}-\frac{2}{n_l}\sum_{i=n_1+\cdots+n_{l-1}+1}^{n_l} \hat
{g}_{l,i}(\mathbf{t}_i).
\nonumber
\end{eqnarray}
%

In general, cross-validation using the leave one out estimator is
computationally inefficient. However, for sufficiently large samples
and $l=1,\ldots,q,m$,
a useful simplification is obtained from the approximation
\[
\int\hat{g}_l(\mathbf{x})g_l(\mathbf{x})\,d\mathbf{x}
\stackrel{n\rightarrow\infty}{\rightarrow}\int\tilde{g}_l(\mathbf
{x})g_l(\mathbf{x})\,d\mathbf{x}.
\]
%
Moreover,
\begin{eqnarray*}
&&
E\biggl[\int\tilde{g}_l(\mathbf{x})g_l(\mathbf{x})\,d\mathbf{x}
\biggr]\\
&&\qquad=E\Biggl[\int\frac{1}{n h_n^p}\sum_{i=1}^m\sum_{j=1}^{n_i}\frac
{w_l(\mathbf{x}_{ij})}{\sum_{k=1}^m \zeta_k w_k(\mathbf{x}_{ij})}
K\biggl(\frac{\mathbf{x}-\mathbf{x}_{ij}}{h_n}\biggr)g_l(\mathbf
{x})\,d\mathbf{x}\Biggr]\\
&&\qquad=h_n^{-p}\int\int K\biggl(\frac{\mathbf{x}-\mathbf{y}}{h_n}
\biggr)g_l(\mathbf{x})g_l(\mathbf{y})\,d\mathbf{x}\,d\mathbf{y}\\
&&\qquad=E\biggl[h_n^{-p}K\biggl(\frac{\mathbf{x}-\mathbf{y}}{h_n}
\biggr)\biggr]\\
&&\qquad=E\biggl[\frac{1}{n_l(n_l-1)h_n^p}\sum_{i\neq j}K\biggl(\frac
{\mathbf
{x}_{li}-\mathbf{x}_{lj}}{h_n}\biggr)\biggr].
\end{eqnarray*}
Thus, for sufficient large $n$, an unbiased estimator for
$\int\tilde{g}_l(\mathbf{x})g_l(\mathbf{x})\,d\mathbf{x}$ is
\[
\frac{1}{n_l(n_l-1)h_n^p}\sum_{i\neq j}K\biggl(\frac{\mathbf
{x}_{li}-\mathbf{x}_{lj}}{h_n}\biggr).
\]
Therefore, an alternative way to find $h_n$ is by minimizing
%
\begin{eqnarray}\label{leaveoneout3}
&&
h_n^{-p}\sum_{i=1}^n \sum_{i'=1}^n \hat{p}(\mathbf{t}_i) \hat
{w}_l(\mathbf{t}_i)\hat{p}(\mathbf{t}_{i'})\hat{w}_l(\mathbf
{t}_{i'})\int K(\mathbf{z})K\biggl(\mathbf{z}+\frac{\mathbf
{t}_i-\mathbf
{t}_{i'}}{h_n}\biggr)\,d\mathbf{z}\nonumber\\[-8pt]\\[-8pt]
&&\qquad{}-\frac{2}{n_l(n_l-1)h_n^p}\sum_{i\neq j}K\biggl(\frac{\mathbf
{x}_{li}-\mathbf{x}_{lj}}{h_n}\biggr).
\nonumber
\end{eqnarray}

Cross-validation has the advantage that (\ref{leaveoneout1})
and (\ref{leaveoneout3}) can easily be modified if we wish to use
different bandwidths $h_1, \ldots, h_p$ to smooth each term, respectively.

\section{Semiparametric regression}
\label{Semiparametricregression}

Suppose we have $m=q+1$ data sets or samples of $p$-dimensional
vectors, where each vector consists of $p-1$ covariates
and one response, and assume that the $i$th sample size is $n_i$.
Thus, for $i=1,\ldots,q,m, j=1,\ldots,n_i$, we have
\[
\bigl(x_{ij1}, x_{ij2}, \ldots, x_{ij(p-1)}, y_{ij} \bigr)
\sim g_{i}\bigl(x_1,\ldots,x_{(p-1)},y\bigr).
\]
We choose $g\equiv g_{m}(x_1,\ldots,x_{(p-1)},y)$ as a reference or baseline
probability density function (p.d.f.), and let each
$g_{i}(x_1,\ldots,x_{(p-1)},y), i=1,\ldots, q$,
be an exponential distortion or tilt of the reference distribution,
%
\begin{equation}\label{LDDensityRatioModel}
\frac{g_{i}(\mathbf{x})}{g(\mathbf{x})}=
\exp(\alpha_{i}+\bfbeta_{i}'\mathbf{x}),\qquad i=1,\ldots,q,
\end{equation}
where $\mathbf{x}=(x_1,\ldots,x_{(p-1)},y)^{\prime}$ and
$\bfbeta_{i}=(\beta_{i1},\ldots,\beta_{ip} )^{\prime}$. Since the
$g_{i}(\mathbf{x}), i=1,\ldots,q,m$, are probability densities,
$\bfbeta_{i}=\mathbf{0}$ implies $\alpha_{i}=0$, $ j=1,\ldots,q$. It
follows that the hypothesis $H_0\dvtx
\bfbeta_1=\cdots=\bfbeta_q=\mathbf{0}$ implies equidistribution: all
the $g_i$ are equal.
%
\begin{remark}
Model (\ref{LDDensityRatioModel}) is motivated from
the ratio of two multivariate normal densities assuming the same
covariance matrices
[\citet{And71}, \citet{Kedetal09}].
It is a special case of model
(\ref{SDRM1}) with
$w(\mathbf{x},\bftheta_i)=w(\mathbf{x},\alpha_i, \bfbeta_i)
\equiv\exp(\alpha_{i}+\bfbeta_{i}'\mathbf{x})$.
\end{remark}

Let $\mathbf{t}$ denote the vector of combined data of length
$n = n_1+n_2+\cdots+n_m$. Following the method of constrained empirical
likelihood, we obtain
score equations for $\hat{\alpha}_{j}$ and $\hat{\bfbeta}_{j}$:
%
\begin{eqnarray}
\label{Scoreeq1}\qquad
\frac{\partial{l}}{\partial\alpha_j} &=&
-\sum_{i=1}^{n}\frac{\rho_{j}w_{j}(\mathbf{t}_i)}
{1+\rho_1w_1(\mathbf{t}_i)+
\cdots+\rho_{q}w_q(\mathbf{t}_i)}+n_j=0, \\
\label{Scoreeq2}
\frac{\partial{l}}{\partial\bfbeta_{j}} &=&
-\sum_{i=1}^{n}\frac{\rho_{j}w_{j}(\mathbf{t}_i)\mathbf{t}_i}{1+
\rho_1w_1(\mathbf{t}_i)+
\cdots
+\rho_{q}w_q(\mathbf{t}_i)}+\sum_{i=1}^{n_j}
(x_{ji1},\ldots,y_{ji})' =0
\end{eqnarray}
for $j=1,\ldots,q$ and $\rho_j =n_j/n_m$. Then
%
\begin{eqnarray}
\label{Estimatedpi}
\hat{p}_i&=&\frac{1}{n_m}\cdot\frac{1}
{1+\rho_1 \hat{w}_1(\mathbf{t}_i)+ \cdots
+\rho_{q}\hat{w}_q(\mathbf{t}_i)},
\\
\label{EstimatedMultivariateG}
\hat{G}(\mathbf{t})&=&\frac{1}{n_m}\cdot\sum_{i=1}^{n}
\frac{I(\mathbf{t}_i\leq\mathbf{t})}
{1+\rho_1\hat{w}_1(\mathbf{t}_i)+ \cdots
+\rho_{q}\hat{w}_q(\mathbf{t}_i)},
\end{eqnarray}
where
$(\mathbf{t}_i\leq\mathbf{t})$ is defined componentwise,
$\hat{w}_{j}(\mathbf{t}_{i})=\exp(\hat{\alpha}_{j}+\hat{\bfbeta}_{j}'
\mathbf{t}_{i})$, and
$I(B)$ is the indicator of the event $B$.
Following \citet{Lu07},
we can show that
as $n\rightarrow\infty$ the estimators $\hat{\bftheta}=(\hat
{\alpha
}_{1},\ldots, \hat{\alpha}_{q},
\hat{\bfbeta}_{1},\ldots,\hat{\bfbeta}_{q})'$ are asymptotically normal.


\subsection{Computing $E[y|\mathbf{x}]$ using the density ratio
model}\label{EygivenxSDRM}

Under the $p$-dimen\-sional density ratio model we can predict the
response $y$ given the covariate information
$x_1,x_2,\ldots,x_{(p-1)}$ for any of the $m$ data sets as follows:
%
\begin{eqnarray}\label{CondExp}
\hat{E_j}\bigl(y \mid x_1,\ldots,x_{(p-1)}\bigr)=
\sum_i^{n_j} y_i \frac{\hat{g}_j(x_1,\ldots,x_{(p-1)},y_i)}
{\sum_{y_i}\hat{g}_j(x_1,\ldots,x_{(p-1)},y_i)},\nonumber\\[-8pt]\\[-8pt]
&&\eqntext{j=1,\ldots,q,m.}
\end{eqnarray}
The $\hat{g}_j$ in (\ref{CondExp}) are the semiparametric kernel
density estimates.
Theorem~5 in \citet{VouKedGra} establishes the
consistency of~(\ref{CondExp}) under some conditions.

It is interesting to compare the semiparametric estimator for
$E[y|\mathbf{x}]$ against the Nadaraya--Watson estimator [\citet{Nad64}, \citet{Wat64}]
and the estimators obtained from linear
regression [\citet{Ren00}],
and GAM [\citet{HasTib90}, \citet{Woo06}].

\subsection{Diagnostic plots and measures of goodness of fit}
\label{measuresofgoodnessoffit}

The density ratio model motivates graphical and quantitative diagnostic
tools for measuring both goodness of fit of the model
and the quality of the regression (\ref{CondExp}). Goodness-of-fit
tests have been proposed by \citet{Gil04}, \citet{QinZha97}
and Zhang (\citeyear{Zha99}, \citeyear{Zha01}, \citeyear{Zha02}),
where the appropriateness of the model is
judged by the closeness of the estimated reference distribution to the
corresponding empirical distribution. \mbox{\citet{Bon07}}
suggests a
reformulation of this in terms of the corresponding kernel density
estimates. We suggest data analytic tools to measure
discrepancies stemming from all case \textit{and} control (reference) groups.

Graphical evidence of goodness of fit can be obtained from the plots
of~$\hat{G}_i$ versus the corresponding empirical multivariate
distribution function $\tilde{G}_i$, $i=1,\ldots,q,m$, evaluated at
some selected $p$-dimensional points as to obtain two-dimensional
plots. Figures~\ref{hatGvstildeG12} and~\ref{hatGvstildeG34} in the
next section are examples of this. We refer to these plots as
diagnostic plots.

We found the following measure of goodness of fit useful. Consider the
$i$th sample of size $n_i$. Let $x_\alpha$ be the number of times the
estimated semiparametric cdf falls in the estimated $1-\alpha$
confidence interval obtained from the corresponding empirical
cdf, both evaluated at the sample points. Define
%
\begin{equation}\label{R2alpha}
R_{\alpha,k}^2=1-\exp\biggl\{-\biggl(\frac{x_\alpha}{n_i-x_\alpha
}\biggr)^k\biggr\},
\end{equation}
where $k>0$, and $k$ and $\alpha$ are free parameters, which can be set
by the user. Observe that:
\begin{itemize}
\item
$R_{\alpha,k}^2$ takes values between $0$ and $1$,
being close to $1$ when $x_\alpha$ approaches $n_i$ and close to $0$
when $x_\alpha$
is close to $0$.
\item
$R_{\alpha,k}^2$ is a flexible criterion that can be adjusted by
changing the parameters~$\alpha$ and $k$. Larger $\alpha$ means smaller
confidence interval bounds.
\item
Computing $R_{\alpha,k}^2$ is both simple and fast.
\end{itemize}

We now describe three natural alternatives to $R_{\alpha,k}^2$. First,
as in multiple regression, goodness of fit may be approached by
residual analysis. In this vein, we define ``$R^2$'' as in linear regression:
%
\begin{equation}\label{R21}
R_1^2=\frac{\sum(\hat{y}_i-\bar{y})^2}{\sum(y_i-\bar{y})^2}.
\end{equation}
Next, define
%
\begin{equation}\label{R22}
R_2^2 =\operatorname{corr}(y,\hat{y})^2.
\end{equation}
Notice that $R_1^2$ and $R_2^2$ depend on $\hat{y}$, and the process
of calculating $\hat{y}$ involves selecting the bandwidth, making the
process of calculating them complicated.
In addition, some early simulation results suggested that they are
misleading as measures of goodness of fit, and, thus,
they were rejected.

Next, following \citet{QinZha97}, define
%
\begin{equation}\label{R23}
R_3^2=\exp\bigl(-\sqrt{n}\cdot{\max}|\tilde{G}_i-\hat{G}_i |\bigr).
\end{equation}
Clearly, $R_3^2$ takes values between $0$ and $1$. Alternatives to
$R_3^2$ are
$\exp(-\sqrt{n}\cdot\mathrm{median} |\tilde{G}_i-\hat{G}_i |)$
or
$\exp(-\frac{1}{n}\sum|\tilde{G}_i-\hat{G}_i | ^2)$.

The following simulation
study compares $R_{\alpha,k}^2$ and $R_3^2$. The simulation
suggests that $R_{\alpha,k}^2$ is a useful indicator of
goodness of fit.


\section{Some simulation results}
\label{SomeSimulationResults}

In the present simulation study $m=2$, $g_2$
denotes the reference distribution, and the results were obtained from
100 runs
(repetitions) of the following four bivariate cases:
\begin{longlist}[(4)]
\item[(1)]
$g_1\!\sim\! N((0,0)',\bfSigma))$, $g_2\!\sim\!N((0,0)',\bfSigma))$
with $\bfSigma\!=\!\bigl({4 \atop 2} \enskip {2 \atop 3}\bigr)$,
$n_1\!=\!40$, \mbox{$n_2\!=\!30$}.
\item[(2)]
$g_1\!\sim\!N((0,0)',\bfSigma))$, $g_2\!\sim\!N((1,1)',\bfSigma))$
with $\bfSigma\!=\!\bigl({3 \atop 1} \enskip {1 \atop 2}\bigr)$,
$n_1\!=\!200$, \mbox{$n_2\!=\!200$}.
\item[(3)]
$g_1$ from standard two-dimensional Multivariate Cauchy and
$g_2$ from two-dimensional Multivariate Cauchy with
$\bfmu=(1,1)'$, $\mathbf{V}=\bigl(
{5 \atop 5} \enskip {5 \atop 10} \bigr)$,
\mbox{$n_1=200$}, \mbox{$n_2=200$}.
\item[(4)]
$g_1$ from standard two-dimensional Multivariate Cauchy and
$g_2$ from uniform distribution on the triangle
$(0,0),(6,0),(-3,4)$, and
$n_1\!=\!200$, \mbox{$n_2\!=\!200$}.
\end{longlist}

The normal distribution follows the density ratio model,
but this is not true for the Cauchy and the uniform distributions.
Hence, we expect to see straight lines in the diagnostic plots
and high $R^2$'s, as defined above, in
cases~($1$) and ($2$). On the other hand, we expect to see deviations
from straight lines
in the diagnostic plots and lower $R^2$'s in cases ($3$) and ($4$).

\begin{figure}

\includegraphics{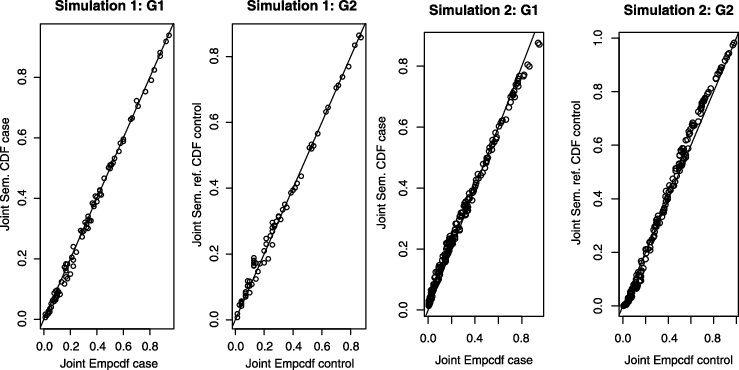}

\caption{Case-control plots of $\hat{G}_i$ vs. $\tilde{G}_i, i=1,2$,
simulations (1) and (2).}
\label{hatGvstildeG12}\vspace*{-3pt}
\end{figure}

\begin{figure}[b]

\includegraphics{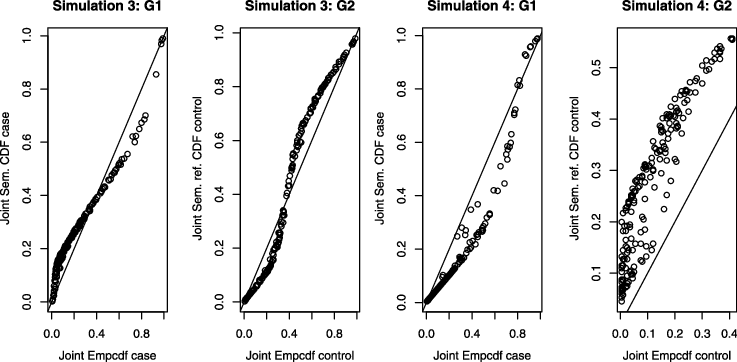}

\caption{Case-control plots of $\hat{G}_i$ vs. $\tilde{G}_i, i=1,2$,
simulations (3) and (4).}
\label{hatGvstildeG34}
\end{figure}

Figures~\ref{hatGvstildeG12} and~\ref{hatGvstildeG34} show\vspace*{1pt}
the estimated $\hat{G}_1$ and $\hat{G}_2$ [where $\hat{G}_1$ is the
exponential tilt of $\hat{G}_2$ defined in
(\ref{EstimatedMultivariateG})] versus the empirical cdf $\tilde{G}_1$
and $\tilde{G}_2$, respectively, all obtained from one run of the
simulated case-control data, and evaluated at selected points
in~$\mathbb{R}^2$. As expected, in cases (1) and (2), there is almost a
perfect agreement between $\hat{G}_i$ versus $\tilde{G}_i, i=1,2$,
whereas Figure~\ref{hatGvstildeG34} shows clearly that the density
ratio model is not appropriate for the data from cases ($3$) and ($4$).

A comparison of $R_3^2$ and $R_{\alpha,k}^2$
obtained from $100$ runs is given in Table~\ref{Goodnessoffitcomparison2}.
It seems that $R_3^2$ is sensitive to outliers and can give low values
even for data that follow the density ratio model [e.g., case (2)].
On the other hand, the proposed measure $R_{\alpha,k}^2$ classifies
correctly the four cases, giving high values for simulations ($1$)
and\vadjust{\goodbreak}
(2) and low values for (3) and (4). The values of $R_{\alpha,k}^2$ in
Table~\ref{Goodnessoffitcomparison2} were calculated
with $k=2$ and $1-\alpha=95\%$. We observed that, by lowering
$1-\alpha
$, $R_{\alpha,k}^2$ gets closer to $0$ for cases (3) and (4).

\begin{table}
\tablewidth=185pt
\caption{Comparison\vspace*{1pt} of $R_3^2$ and $R_{0.05,2}^2$ for $100$ repetitions
of case and control}\label{Goodnessoffitcomparison2}
\begin{tabular*}{\tablewidth}{@{\extracolsep{\fill}}lccd{1.4}@{}}
\hline
\textbf{Run} &\textbf{Group} & $\bolds{R_3^2}$
& \multicolumn{1}{c@{}}{$\bolds{R_{0.05,2}^2}$} \\
\hline
(1) & Case &0.6307 &1 \\
& Control &0.5976 &1 \\
(2) & Case &0.3912 &0.9353 \\
& Control &0.3766 &0.9718\\
(3) & Case &0.1080 &0.3342\\
& Control &0.1129 &0.3324 \\
(4) & Case &0.0507 &0.3361 \\
& Control &0.0495 &0.0033 \\
\hline
\end{tabular*}
\end{table}

\begin{table}
\tablewidth=185pt
\caption{Bandwidth (BW) selection using formula (4) in
Voulgaraki, Kedem and Graubard (\protect\citeyear{VouKedGra})}
\label{CVoptimal}
\begin{tabular*}{\tablewidth}{@{\extracolsep{\fill}}lcc@{}}
\hline
& \textbf{Case} & \textbf{Control} \\
& \textbf{BW} & \textbf{BW} \\
\hline
Simulation 1 &0.46 & 0.47\\
Simulation 2 &0.33 & 0.51\\
\hline
\end{tabular*}
\end{table}

\begin{table}[b]
\caption{Bandwidth (BW) selection using the cross-validation method
(\protect\ref{leaveoneout1})}\label{CVleave1out}
\begin{tabular*}{\tablewidth}{@{\extracolsep{\fill}}lcccccc@{}}
\hline
& \multicolumn{3}{c}{\textbf{Case}} &
\multicolumn{3}{c@{}}{\textbf{Control}} \\[-4pt]
& \multicolumn{3}{c}{\hrulefill} & \multicolumn{3}{c@{}}{\hrulefill}\\
& \multirow{2}{39pt}[-7pt]{\centering{\textbf{Same BW} $\bolds{h}$}} &
\multicolumn{2}{c}{\textbf{Diff. BWs}} &
\multirow{2}{39pt}[-7pt]{\centering{\textbf{Same BW} $\bolds{h}$}}
& \multicolumn{2}{c@{}}{\textbf{Diff. BWs}} \\[-4pt]
& & \multicolumn{2}{c}{\hrulefill} & & \multicolumn{2}{c@{}}{\hrulefill}\\
&  & $\bolds{h_1}$ & $\bolds{h_2}$ & $\bolds{h}$
& $\bolds{h_1}$ & $\bolds{h_2}$\\
\hline
Simulation 1 &0.61 &0.90 &0.40 & 0.59 &0.31 & 0.61\\
Simulation 2 &0.38 &0.50&0.20 & 0.61 &0.36 & 0.71\\
\hline
\end{tabular*}
\end{table}


As noted earlier, calculating
the semiparametric $\hat{E}[Y|X]$ for cases (1) and~(2) entails
bandwidth selection, which can be done either via the asymptotically
optimal formula (4)
in \citet{VouKedGra}, replacing
$g_l$ with $N(\hat{\bfmu}, \hat{\bfSigma})$ (parameters estimated
from the data), or via cross-validation and minimize
either (\ref{leaveoneout1}) or (\ref{leaveoneout3}) (which
also allows different bandwidths
$h_1, \ldots, h_p$ to smooth the different terms).
Tables
\ref{CVoptimal}--\ref{CVasymptotic} summarize the results for
the estimated bandwidths for one run of the simulations,
using equations (4) in \citet{VouKedGra},
(\ref{leaveoneout1})
and~(\ref{leaveoneout3}). The integrals in (4) in \citet{VouKedGra}
were
calculated using \textit{Mathematica}.
There were no significant differences in the regression
results using single or multiple bandwidths.


Using the semiparametric model, the standard normal distribution for the
kernel and (\ref{CondExp}), we estimated $E[Y|X]$ for a single
predictor. Table~\ref{MSEcomparison} provides
MSE and MAE comparisons between the different methods for the first two
simulations. In the table SP stands for semiparametric regression,
MR for multiple regression, GAM for generalized additive model and
NW for Nadaraya--Watson. We did not estimate $E[Y|X]$ for simulations
$3$ and $4$ because the semiparametric model is not applicable in
these cases (and was rejected as we saw from the $R^2$ comparisons). In
simulations $1$--$2$, for both case and control, we fitted a thin
plate regression spline GAM assuming the normal distribution and identity
link. The results for tensor product were almost identical.
In simulation $1$ the GAM line was
almost identical to the multiple regression line. We see that
the semiparametric regression performs comparably with the other
methods in terms of MSE and MAE.

\begin{table}
\caption{Bandwidth (BW) selection using the cross-validation method
(\protect\ref{leaveoneout3})}\label{CVasymptotic}
\begin{tabular*}{\tablewidth}{@{\extracolsep{\fill}}lcccccc@{}}
\hline
& \multicolumn{3}{c}{\textbf{Case}} &
\multicolumn{3}{c@{}}{\textbf{Control}} \\[-4pt]
& \multicolumn{3}{c}{\hrulefill} & \multicolumn{3}{c@{}}{\hrulefill}\\
& \multirow{2}{39pt}[-7pt]{\centering{\textbf{Same BW} $\bolds{h}$}} &
\multicolumn{2}{c}{\textbf{Diff. BWs}} &
\multirow{2}{39pt}[-7pt]{\centering{\textbf{Same BW} $\bolds{h}$}}
& \multicolumn{2}{c@{}}{\textbf{Diff. BWs}} \\[-4pt]
& & \multicolumn{2}{c}{\hrulefill} & & \multicolumn{2}{c@{}}{\hrulefill}\\
&  & $\bolds{h_1}$ & $\bolds{h_2}$ & $\bolds{h}$
& $\bolds{h_1}$ & $\bolds{h_2}$\\
\hline
Simulation 1 &0.64 &0.90 &0.50 & 0.63 &0.21 & 0.71\\
Simulation 2 &0.30 &0.40&0.20 & 0.74 &0.11 & 0.96\\
\hline
\end{tabular*}
\end{table}

\begin{table}[b]
\caption{MAE and MSE comparison between regression methods,
for simulations 1 and 2. $G_1$, $G_2$ signify case and control,
respectively}\label{MSEcomparison}
\begin{tabular*}{\tablewidth}{@{\extracolsep{\fill}}lccccccccc@{}}
\hline
& &\multicolumn{4}{c}{\textbf{MSE}} &
\multicolumn{4}{c@{}}{\textbf{MAE}} \\[-4pt]
&& \multicolumn{4}{c}{\hrulefill} &
\multicolumn{4}{c@{}}{\hrulefill}\\
& & \textbf{SP} & \textbf{MR} & \textbf{GAM}
& \textbf{NW} & \textbf{SP} & \textbf{MR} & \textbf{GAM} &
\textbf{NW} \\
\hline
Simulation 1& $G_1$ &0.913 &0.834 &0.834 &0.851 & 0.752 &0.741 &0.741
&0.736 \\
& $G_2$ &0.856 &0.892 &0.892 &0.849 & 0.750 &0.786 &0.786 &0.740 \\
Simulation 2& $G_1$ &0.820 &0.841 &0.799 &0.792 & 0.723 &0.730 &0.709
&0.704 \\
& $G_2$ &1.740 &1.482 &1.429 &1.388 & 1.001 &0.992 &0.958 &0.946 \\
\hline
\end{tabular*}
\end{table}


%
%

\section{Application to testicular germ cell cancer}
\label{ApplicationtoTesticularGermCellCancer}

Testicular germ cell tumor (TGCT) is a common cancer among U.S. men,
mainly in the age group of 15--35 years [\citet{McGetal03}].
In \citet{McGetal07}
it was shown that
increased risk was significantly related to height, whereas body
mass index was not significant. In \citet{Kedetal09}, using the
two-dimensional semiparametric model, it was shown that
\textit{jointly} height and weight are significant risk factors.
The TGCT data consist of age (years), height (cm)
and weight (kg) of $1691$ individuals, of which $n_1=763$ are cases
and $n_2=928$ belong to the control group. We considered two cases:
the 2D TGCT data set with variables height and weight and the
3D TGCT data set with variables height, weight and age. In both cases
the control distribution was the reference distribution.

Equation (4) in \citet{VouKedGra},
(\ref{leaveoneout1}),
(\ref{leaveoneout3}) with kernel $K=N(\mathbf{0},\mathbf{1})$
and $w(\mathbf{x},\bftheta_i)\equiv
\exp(\alpha_{i}+\bfbeta_{i}'\mathbf{x})$
were used
to calculate the different bandwidths. The three methods gave similar results.
For the 2D TGCT data set, the case bandwidths were
$1.01$ and $3.51$ for height and weight, respectively,
whereas, for control, we used $2.02$ and $1.01$. For the 3D TGCT
the bandwidths were $2.24$
for control and $2.5$ for case.

\begin{figure}

\includegraphics{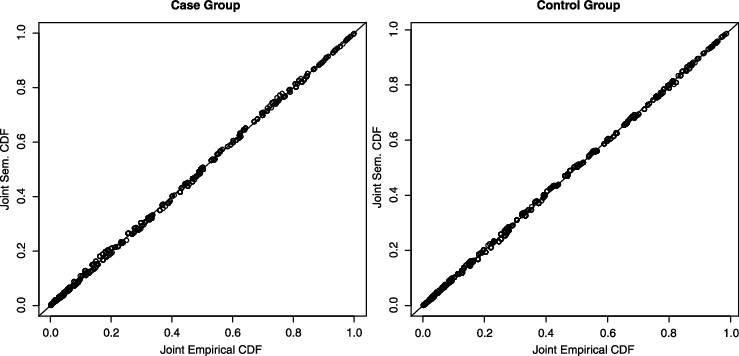}

\caption{2D problem: diagnostic plots of $\hat{G}_i$ versus
$\tilde{G}_i, i=1,2$, evaluated at (height, weight) pairs
for the case and control groups from the TGCT
data.}
\label{hatGvstildeG2DTGCT}
\end{figure}

\begin{figure}

\includegraphics{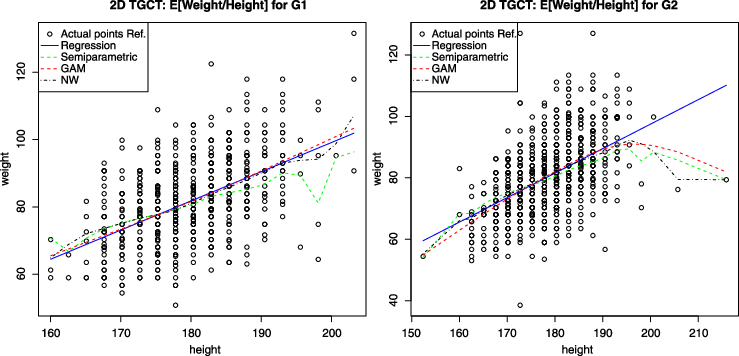}

\caption{Comparison of $\hat{E}[\mbox{weight}\mid \mbox{height}]$ for the 2D TGCT
data set.}
\label{Eweightheight2DTGCT}
\end{figure}

\begin{figure}[b]

\includegraphics{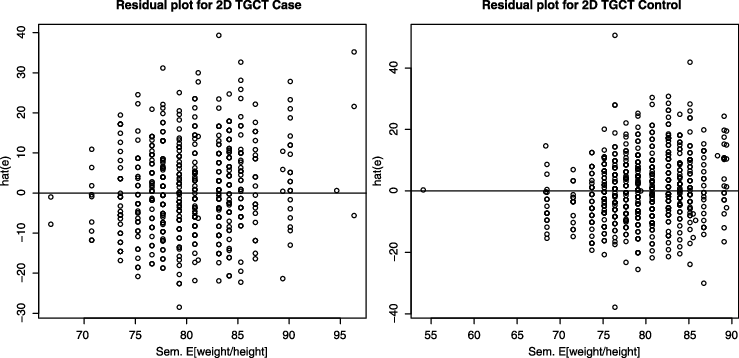}

\caption{Residual plots for the semiparametric model in the 2D TGCT
data set.}
\label{Residuals2DTGCT}
\end{figure}

Before applying the three-dimensional density ratio model to the TGCT
data, it is interesting to apply the two-dimensional model to get
a prediction of weight given height only. As Figure
\ref{hatGvstildeG2DTGCT} shows, the density
ratio model is a suitable model for the TGCT data: there is almost a
perfect agreement between the plots of the semiparametric
$\hat{G}_i$ and the corresponding empirical~$\tilde{G}_i$, $i=1,2$.
The value of $R^2_{0.20,1}$ is $1$ for both case and control.
Figure~\ref{Eweightheight2DTGCT} shows the estimated $E[Y|X]$ using
(\ref{CondExp}) for the case and control groups, where in the 2D TGCT
data set $Y$ is weight and $X$ is height. Superimposed are
the regression lines obtained from linear regression under the normal
assumption, GAM and the Nadaraya--Watson regression.
For the 2D TGCT data, assuming normal distribution and identity link,
we fitted a tensor product GAM; there were essentially no
differences between the different kinds of splines.
We notice that all models give similar results.
The residual plots for the semiparametric model in Figure
\ref{Residuals2DTGCT} are centered around zero.

\begin{figure}

\includegraphics{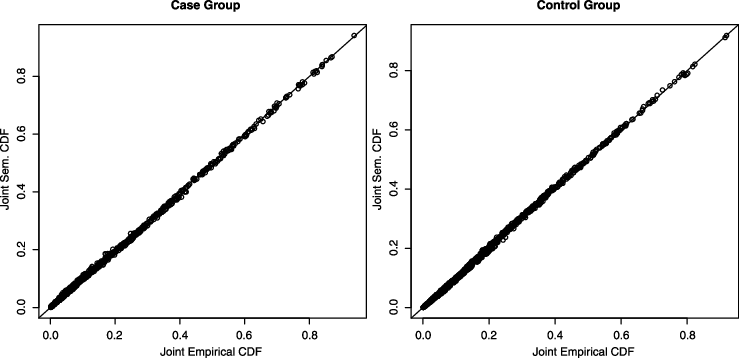}

\caption{Case-control diagnostic plots of $\hat{G}_i$ versus $\tilde
{G}_i, i=1,2$,
for the 3D TGCT problem: the $\hat{G}_i, \tilde{G}_i$ are evaluated at
selected (age, height, weight) triplets.}
\label{hatGvstildeG3DTGCT}
\end{figure}

Next we fitted the 3D TGCT data with variables age, height and
weight. The semiparametric model is an appropriate model for this data
set as Figure~\ref{hatGvstildeG3DTGCT} shows. The value of
$R^2_{0.20,1}$ is $1$ for both case and control. \textit{An advantage of the
method is that it gives estimates for the joint probabilities of
age, height and weight in both case and control groups as in
Table~\ref{JointProbabilities2DTGCT}}. The table shows
the two groups are moderately different.

\begin{table}[b]
\tablewidth=320pt
\caption{Some joint probabilities of age, height and weight in the
case and
control groups}\label{JointProbabilities2DTGCT}
\begin{tabular*}{\tablewidth}{@{\extracolsep{\fill}}l c c @{}}
\hline
\textbf{Probability} & \textbf{Case} & \textbf{Control} \\
\hline
Pr(A $\leq$ 45, H $\leq$ 152.40, W $\leq$ 58.967) &0.000378 &0.000767
\\
Pr(A $\leq$ 26, H $\leq$ 165.10, W $\leq$ 58.967) & 0.004502 &0.007074
\\
Pr(A $\leq$ 29, H $\leq$ 177.80, W $\leq$ 65.317) & 0.042723 &
0.054313 \\
Pr(A $\leq$ 33, H $\leq$ 185.42, W $\leq$ 70.307) & 0.157968 &
0.184774 \\
Pr(A $\leq$ 34, H $\leq$ 180.34, W $\leq$ 79.832) & 0.316077 &
0.362967 \\
Pr(A $\leq$ 37, H $\leq$ 180.34, W $\leq$ 89.811) & 0.513664 &
0.575512 \\
Pr(A $\leq$ 40, H $\leq$ 187.96, W $\leq$ 94.801) & 0.797157 &0.833803
\\
Pr(A $\leq$ 43, H $\leq$ 200.66, W $\leq$ 99.790) & 0.943058 &
0.956300 \\
Pr(A $\leq$ 45, H $\leq$ 203.20, W $\leq$ 117.934) & 0.995010 &
0.996560\\
\hline
\end{tabular*}
\end{table}

\begin{figure}[b]
\vspace*{-3pt}
\includegraphics{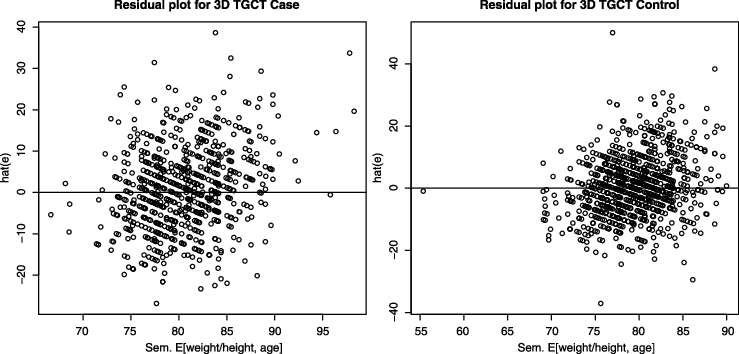}

\caption{Residual plots for the semiparametric model in the 3D TGCT
data set.}
\label{Residuals3DTGCT}
\end{figure}

In order to calculate $\hat{E}[Y|\mathbf{X}]$ for the case and control
groups, we used~(\ref{CondExp}), where in the 3D TGCT data
set $Y$ is weight and $\mathbf{X}$ represents jointly height and age.
Figure~\ref{Residuals3DTGCT} shows the residual plots for the
semiparametric model. Table~\ref{MSEcomparisonTGCTdata} gives the
MSE and MAE comparison between the different regression methods
for the 2D and the 3D TGCT data. For the 3D TGCT data, assuming normal
distribution and identity link, we fitted a thin plate regression
spline GAM because it produced better looking residual and Q--Q plots.
The semiparametric regression performs comparably with the other
estimators, although it has a somewhat higher MSE.
These results can be explained
by the fact that our method consists of an extra step of density
estimation. However, we have the extra advantage that we also obtain
joint probabilities of the variables,
unlike multiple regression and GAM.

\begin{table}
\caption{MAE and MSE comparison of the semiparametric,
multiple, GAM and Nadaraya--Watson regression methods
for 2D and 3D TGCT data}\label{MSEcomparisonTGCTdata}
\begin{tabular*}{\tablewidth}{@{\extracolsep{\fill}}lcd{3.3}ccccccc@{}}
\hline
& &\multicolumn{4}{c}{\textbf{MSE}} &
\multicolumn{4}{c@{}}{\textbf{MAE}} \\[-4pt]
&& \multicolumn{4}{c}{\hrulefill} &
\multicolumn{4}{c@{}}{\hrulefill}\\
& & \multicolumn{1}{c}{\textbf{SP}} & \textbf{MR} & \textbf{GAM}
& \textbf{NW} & \textbf{SP} & \textbf{MR} & \textbf{GAM} &
\textbf{NW} \\
\hline
2D TGCT & $G_1$ &104.003 &99.510 &99.250 &98.648 & 7.947 &7.784 &7.770
&7.774 \\
& $G_2$ &93.010 &92.264 &90.284 &90.332 & 7.347 &7.296 &7.246 &7.241
\\
3D TGCT & $G_1$ &98.283 &96.367 &96.091 &89.124 & 7.770 &7.679 &7.672
&7.390 \\
& $G_2$ &91.643 &90.291 &88.147 &86.932 & 7.280 &7.244 &7.173 &7.139
\\
\hline
\end{tabular*}    \vspace*{-3pt}
\end{table}

Tables~\ref{Controlexpectedvalues} and~\ref{Caseexpectedvalues}
give some predicted values for weight given age and
height for the two groups. The results from the different
methods are not much different.


We end this section by providing $\hat{E}(y | \mathbf{x})$ in (\ref
{CondExp})
to help the reader interpret the results of the semiparametric analysis.
Tables~\ref{SemCaseControlpred1} and
\ref{SemCaseControlpred2} give the case-control weight
predictions (\ref{CondExp}) and
the actual weights.
From the tables, as expected,
$\hat{E}(y | \mathbf{x})$ in
(\ref{CondExp}) tends to be close to the average of $y$'s
which correspond to the same $\mathbf{x}$.
Empty entries in the table correspond to subjects
with the same height and age (i.e., same $\mathbf{x}$),
but possibly different weights. The averaging property can be seen
by averaging the run of weights in the ``empty cells'' and the run ``upper
point.'' Thus, for example,
the control-weights corresponding to age $22$ and height $175.26$
average to $74.3894$ and the prediction is 76.62195.
Across different ages, except for heights less than 167.64 cm, the
estimated conditional expectation in
cases consistently has greater body weights than controls, indicating
that later life exposures such as increased caloric diet intake and/or
reduced energy expenditure and lack of physical exercise may increase
the risk of testicular cancer.\vspace*{-2pt}

\section{Summary}

In this paper we have shown that using our proposed semi-parametric
regression method we can detect an important increased risk of germ
cell testicular cancer with greater body weight after adjusting for
age and height that was not found with these same data using standard
logistic regression modeling. This is an important finding because
body weight is likely a later life exposure involving dietary caloric
intake and/or energy expenditure from physical activity. This is in
contrast to height that is influenced by\vadjust{\goodbreak} early life factors such as
genetics, early life nutrition or endogenous or exogenous hormones.
The possibility of intervening to reduce body weight among young men
could help to stem the rise in incidence of testicular cancer.

%
\begin{table}
\caption{Predicted control values of weight given height and age}
\label{Controlexpectedvalues}
\begin{tabular*}{\tablewidth}{@{\extracolsep{\fill}}lcd{3.3}cccc@{}}
\hline
\multicolumn{7}{@{}c@{}}{\textbf{Case}} \\
\hline
\textbf{Age} & \multicolumn{1}{c}{\textbf{Height}}
& \multicolumn{1}{c}{\textbf{Weight}} & \multicolumn{1}{c}{\textbf{SP}}
& \multicolumn{1}{c}{\textbf{MR}} & \multicolumn{1}{c}{\textbf{GAM}}
& \multicolumn{1}{c@{}}{\textbf{NW}} \\
\hline
26 & 193.04 &102.058 &89.81775 &92.47554 &92.80697 &95.96000 \\
24 &167.64 &72.575 &73.59282 &70.00329&70.68805 &71.90371 \\
29&180.34 &65.771 &81.41551 &82.42360&82.17237 &81.60395 \\
38&185.42 &81.647 &86.29762 &89.46406&89.50287 &89.70666 \\
34&195.58 &89.811&89.03635 &97.03194 &98.08814 &92.45555\\
27 &162.56 &58.967 &68.53652 &66.51540 &67.76775 &65.18988\\
\hline
\end{tabular*}
\end{table}
%

%
\begin{table}[b]
\caption{Predicted case values of weight given height and age}
\label{Caseexpectedvalues}
\begin{tabular*}{\tablewidth}{@{\extracolsep{\fill}}lcd{3.3}d{2.5}cd{2.5}c@{}}
\hline
\multicolumn{7}{@{}c@{}}{\textbf{Control}} \\
\hline
\textbf{Age} & \multicolumn{1}{c}{\textbf{Height}} & \multicolumn{1}{c}{\textbf{Weight}}
& \multicolumn{1}{c}{\textbf{SP}} & \multicolumn{1}{c}{\textbf{MR}}
& \multicolumn{1}{c}{\textbf{GAM}} & \textbf{NW} \\
\hline
29 & 180.34 & 90.718 & 81.11841 &82.06293 &83.06542 & 82.35544 \\
39 & 175.26 & 77.111 &79.40282 &80.36549 &79.78087 &80.05940 \\
19 & 172.72 & 63.503 &74.76493 &73.58821 &72.76199 &73.40060\\
33 & 177.80 & 83.915 &80.51759 &80.97707 &81.4916 &81.14195 \\
31 & 190.50 & 102.058 &86.0598 &90.67494 &90.69862 &87.47080 \\
25 & 165.10 & 58.967 & 72.08147 & 68.90777 &68.0279 &69.49050 \\
\hline
\end{tabular*}
\end{table}

The semiparametric regression approach taken in this paper requires
first efficient estimation of multivariate distributions. This can be
achieved under the multidimensional density ratio model, given multiple
data sources of multivariate data, and known tilts up to unknown
parameters. Subject
to this construct, the method produces more efficient kernel
density estimators than the traditional single-sample
kernel density estimator. This is so since
all the finite and infinite-dimensional parameters are estimated from
the entire combined data from all sources, and not just from single
sources. As in the traditional kernel estimation, our kernel estimates
require bandwidths and we have discussed ways for
obtaining optimal and nearly optimal kernel bandwidths.
The process
of fitting the density ratio model and obtaining estimates is quite
straightforward and quick. In this regard, several diagnostic measures
have been suggested.

%
\begin{table}[t!]
\caption{Case-control weight and $\hat{E}[\mbox{weight}|\mbox{height,
age}]$. Empty entries in the table correspond to subjects
with the same height and age, but possibly different weights}
\label{SemCaseControlpred1}
\begin{tabular*}{\tablewidth}{@{\extracolsep{\fill}}lcccd{3.3}d{2.5}@{}}
\hline
& & \multicolumn{2}{c}{\textbf{Control}}
& \multicolumn{2}{c@{}}{\textbf{Case}} \\[-4pt]
& & \multicolumn{2}{c}{\hrulefill}
& \multicolumn{2}{c@{}}{\hrulefill} \\
\textbf{Age} & \multicolumn{1}{c}{\textbf{Height}}
& \multicolumn{1}{c}{\textbf{Weight}}
& \multicolumn{1}{c}{$\bolds{\hat{E}[W\mid H,A]}$}
& \multicolumn{1}{c}{\textbf{Weight}} &
\multicolumn{1}{c@{}}{$\bolds{\hat{E}[W \mid H,A]}$} \\
\hline
27 &162.56 &58.967 &69.08335 & 58.967& 68.53652 \\
28 &162.56 &77.111 &69.05132 & 65.771&68.59858 \\
& &68.039 & & &  \\
30 &165.10 &68.039 &72.20524 & 72.575&72.0028 \\
37 &165.10 &69.40 &72.42138 & 63.503&71.8504 \\
25 &167.64 &86.183 &73.68129 & 72.575 &73.69978 \\
& & & & 90.718 & \\
& & & & 63.503 & \\
30 &167.64 &72.575 &74.81333 & 88.451 &74.93543 \\
18 &170.18 &61.235 &73.67032 & 72.575 &73.67518 \\
32 &170.18 &70.307 &76.53351 & 81.647 &76.64543\\
& &63.503 & & &  \\
37 &172.72 &74.843 &77.88598 & 88.451 &77.9417\\
40 &172.72 &70.307 &77.97789 & 90.718 &78.0441 \\
& &77.111 & & &  \\
22 &175.26 &77.111 &76.62195 & 86.183 &76.70862 \\
& &65.771 & & 65.771  \\
& &79.379 & & 86.183  \\
& &83.915 & & &  \\
& &65.771 & & &  \\
25 &175.26 &68.039 &77.14234 & 79.379 &77.21755\\
& &83.915 & & 72.575 & \\
& &74.843 & & 83.915 & \\
& &83.915 & & 74.843 & \\
& &79.379 & & 72.575 & \\
& &86.183 & & 74.843 & \\
& & & & 61.235 & \\
& & & & 61.235 & \\
& & & & 65.771 & \\
& & & & 79.379 & \\
26 &177.80 &79.379 &78.74752 & 77.111 &78.92705 \\
& &81.647 & & 104.326 \\
& &58.967 & & 77.111  \\
& &81.647 & & &  \\
& &79.379 & & &  \\
& &74.843 & & &  \\
& &88.451 & & &  \\
& &68.039 & & &  \\
42 &177.80 &70.307 &80.50100 & 91.626 &80.67493\\
\hline
\end{tabular*}
\end{table}
%

%
\begin{table}
\caption{Case-control weight and $\hat{E}[\mbox{weight}|\mbox{height,
age}]$ continued. Empty entries in the table correspond to subjects
with the same height and age, but possibly different weights}
\label{SemCaseControlpred2}
\begin{tabular*}{\tablewidth}{@{\extracolsep{\fill}}lcd{3.3}cd{3.3}d{2.6}@{}}
\hline
& & \multicolumn{2}{c}{\textbf{Control}}
& \multicolumn{2}{c@{}}{\textbf{Case}} \\[-4pt]
& & \multicolumn{2}{c}{\hrulefill}
& \multicolumn{2}{c@{}}{\hrulefill} \\
\textbf{Age} & \multicolumn{1}{c}{\textbf{Height}}
& \multicolumn{1}{c}{\textbf{Weight}}
& \multicolumn{1}{c}{$\bolds{\hat{E}[W\mid H,A]}$}
& \multicolumn{1}{c}{\textbf{Weight}} &
\multicolumn{1}{c@{}}{$\bolds{\hat{E}[W \mid H,A]}$} \\
\hline
20 &180.34 &79.832 &79.17623 & 84.368 &79.35688\\
& &65.771 & & 68.039 & \\
& &77.111 & & 79.379 & \\
& &79.379 & & 81.647 & \\
& & & & 72.575 & \\
33 &180.34 &79.379 &81.92536 & 77.111 &82.17689\\
& & & & 81.647 & \\
18 &182.88 &77.111 &80.23013 & 68.039 &80.29011\\
41 &182.88 &79.379 &83.65558 & 86.183 &84.06475\\
19 &185.42 &63.503 &81.45580 & 68.039 &82.09186\\
& & & & 94.347 & \\
& & & & 68.039 & \\
21 &185.42 &86.183 &82.46773 & 79.379 &82.78140\\
& &72.575 & & 77.111 & \\
& &102.058 & & 97.522 & \\
22 &190.50 &97.522 &85.23493 & 86.183 &85.64845\\
& &95.254 & & 71.668 & \\
31 &190.50 &102.058 &86.05980 & 104.326&86.27744\\
& & & & 74.843 & \\
22 &193.04 &86.183 &86.73352 & 102.058&87.18440\\
& & & & 80.739 & \\
24 &193.04 &99.337 &87.50020 & 108.862&88.23938\\
& &86.183 & & &  \\
& &99.790 & & &  \\
& &108.862 & & & \\
34 &193.04 &113.398 &87.72937 & 88.451 &88.58960\\
& & & & 117.934& \\
34 &195.58 &83.915 &88.81524 & 89.811 &89.036535\\
\hline
\end{tabular*}
\end{table}

Going a step further, the estimated distributions can be used in
estimating joint probabilities, in ANOVA-like problems of
determining differences between groups, and
in estimating the conditional expectation of
a response variable given random covariates, provided that
multiple data sources
are available.
An application to predicting weight from height and age
in a case-control problem shows the method competes well with several
well-known regression methods, and at the same time it provides
estimates of
joint probabilities.
Our experience suggests that the method is effective for a small number of
covariates. Computational problems can arise as the number of
variables increases.

We have made some numerical comparisons with GAM, but a general
comparison is not our focus or intention in the present paper.
Still, a few points are in order.
From our limited comparison
it seems the two methods produce similar regression estimates,
and both methods are more complex than multiple regression.
The complexity of GAM stems from their iterative nature, which is
reminiscent of fixed point problems in repeated parametric filtering
where estimates are evaluated at estimates iteratively,
and this may affect the interpretability of
the results [\citet{LiSon02}].
It seems to us that
the semiparametric approach, on the other hand, is somewhat more
straightforward.
We have illustrated in the TGCT data analysis that the resulting
semiparametric regression estimate is indeed close to the average
of the response
conditional on fixed covariates, as one would expect.
This property is
shared by GAM as well. GAM assume additivity.
On the other hand, the
density ratio approach requires an assumption about the tilts.
The suggested diagnostic tools shed light, albeit indirectly, on
the appropriateness of the tilts.

\begin{appendix}
\section*{Appendix}

This Appendix contains supplemental material described in
\citet{VouKedGra}. It provides formal statements and indication of
proofs of the results described in Sections~\ref{Asymptoticresults}
and~\ref{comparisonclassicalsemiparametricestimator}.
\end{appendix}

\section*{Acknowledgments}

The authors wish to thank the referees and the Area Editor
for their dedication, effort and important suggestions.

\begin{supplement}
\stitle{Supplement to ``Semiparametric regression in testicular germ cell
data''}
\slink[doi]{10.1214/12-AOAS552SUPP} 
\slink[url]{http://lib.stat.cmu.edu/aoas/552/supplement.pdf}
\sdatatype{.pdf}
\sdescription{The supplementary material contains all the mathematical proofs of the
lemmas, corrolaries and theorems supporting the statements and results,
including some additional references.}
\end{supplement}

%

\printaddresses

\end{document}